# The Effect of Explicit Polar Species on Conformational Behavior of a Single Polyelectrolyte Chain

Yulia D. Gordievskaya*, Elena Yu. Kramarenko, Alexey A. Gavrilov

*Physics Department, Lomonosov Moscow State University, Moscow 119991, Russian Federation*

*e-mail: gordievskaya@polly.phys.msu.ru*

**Abstract**

In this work we investigated the question of how the molecular nature of the dielectric media and the polymer-solvent dielectric mismatch affect the collapse of a polyelectrolyte chain in solution by means of dissipative particle dynamics simulations. First, we studied whether the explicit treatment of dielectric media as polar beads instead of homogeneous dielectric background results in a different system behavior. We showed that the explicit treatment of polar beads facilitates the chain collapse, i.e. it occurs at smaller values of the electrostatic strength parameter values $\lambda^2 \sim \frac{Q^2}{\varepsilon k_b T}$. We believe that the main reason for such behavior is that the dielectric response is in fact a collective effect, and the "effective" dielectric permittivity is different from the bulk value when the charges are close to each other and/or the density of the charges is high enough. This implies that the value of λ does not have a universal meaning due to the small-scale effects related to the presence of polar species; therefore, changing the total unit charge $Q$ or the temperature $kT$ is not equivalent to changing ε. Next, we investigated how the difference of the dielectric permittivities of the polymer chain and solvent affects the collapse. We showed the polar chain adapts less swollen conformations in the polyelectrolyte regime and collapses easier compared to the case of non-polar chain; the possible reasons for such behavior are discussed.

Understanding the properties of polyelectrolytes – polymers carrying charged groups – is of great importance to industrial and biological systems: being mainly water soluble, they find numerous applications in various fields, in particular, in medicine, food, cosmetic and agricultural industries. Furthermore, biological macromolecules, such as DNA, RNA and proteins, are charged, and their functioning cannot be understood without taking into account the contribution from the electrostatic interactions. The presence of long-range electrostatic interaction makes the behavior of such systems extremely rich and invokes complications in studying the underlying physics.

The electrostatic collapse of a single polyelectrolyte chain in solution is undoubtedly one of the most investigated problems in the field of polyelectrolytes, especially by means of computer simulations and theory.[1–8] When the strength of the electrostatic interactions is relatively low, the counterions are released into the solution, and the chain adopts more swollen conformation compared to a corresponding uncharges chain due to the repulsion between the charged monomer units. Upon increasing the electrostatic interactions strength the counterions start to condense, and, finally, when the majority of counterions are condensed (ionomer regime), the chain collapses to form a dense globule due to the charge-correlation attraction.[4] The collapse of the polyelectrolyte chain is a very important process from a biological point of view: the aforementioned bio-polyelectrolytes, such as RNA or DNA, are densely packed in viruses and cells, and the role of electrostatic interactions in such packaging is likely to be big.[9,10] Moreover, the effective mechanisms of the collapse of a single polyelectrolyte chain are closely related to the aggregation of rigid polyelectrolytes,[11,12] while many biological molecules are in fact rigid polyelectrolytes, and their aggregation plays an important role in the processes occurring on cellular level. Also, the collapse of a single polyelectrolyte chain is closely related to the collapse of polyelectrolyte macro- and microgels, which have many existing and potential applications.[13,14]

Despite the large number of works dedicated to the polyelectrolyte chain collapse, in the vast majority of such works the dielectric permittivity of the medium is treated in a simplified way as a homogeneous background. Such an approach does not take into account the molecular nature of the dielectric response (which is crucial when the distance between the interacting charges becomes comparable to the solvent molecule size) and the actual difference of the dielectric permittivity of a typical polymer chain and solvent. The latter circumstance (often called dielectric mismatch) has been shown to have a dramatic effect on the properties of a wide range of polymer systems.[15–25]

The main idea of this work was to investigate whether the "mean-field" treatment of the dielectric permittivity (i.e. as a homogeneous background), which is often utilized in the simulations of polyelectrolyte systems, reproduces all the important features of the polyelectrolyte chain collapse. To that end, we studied the collapse of a single polyelectrolyte chain using the dissipative particle dynamics simulations with explicit polar species. Moreover, since the effect of the dielectric mismatch has been shown to have dramatic effect on the behavior of polyelectrolyte systems, in this work we also address the question of the collapse of a non-polar polyelectrolyte chain in a polar solvent.

In our work we used the dissipative particle dynamics (DPD) with explicit treatment of electrostatic interaction to perform simulations. DPD is a well known simulation technique which has been utilized to simulate properties of a wide range of polymeric systems. Macromolecules are represented in terms of the bead-and-spring model, with beads interacting by a conservative force (repulsion) $F_{ij}^c$, a bond stretching force (only for connected beads) $F_{ij}^b$, a dissipative force (friction) $F_{ij}^d$, and a random force (heat generator) $F_{ij}^r$. For the dissipative and random force we used the parameters described in the work[26]; a detailed description of the standard DPD approach can be found in the work[26] as well. In order to take into account the electrostatic interactions, we use the method described in the work[27]. The electrostatic force between two charged beads (or sub-beads, see below) is calculated using the following expression:

$$F_{ij}^e = \frac{q_i q_j}{4\pi\varepsilon\varepsilon_0} \begin{cases} \frac{\mathbf{r}_{ij}}{r_{ij}^3} \sin^6\left(\frac{2\pi r_{ij}}{4D}\right), & r_{ij} < D \\ \frac{\mathbf{r}_{ij}}{r_{ij}^3}, & r_{ij} \geq D \end{cases},$$

where $D$ is the damping distance. This approach allows one to prevent overlapping of oppositely charged beads while keeping the exact form of the Coulomb potential at distances larger than $D$; the parameter $D$ is essentially the effective bead size, and the electrostatic interactions at smaller distances are not important for the system behavior. We used $D$=0.65 which was shown[27] to be a good choice for the number density of 3, which was used in the current work.

In order to take explicit polarity of the species into account, we used a recently reported modification[28–30] for DPD. Within that approach, each standard DPD beads is replaced by a "dumb-bell" consisting of two force centers (sub-bead) kept at a fixed small distance $d$ from each other; each sub-bead can carry charge, thus making it possible to simulate beads with dipole moments (see Fig.1). The main advantage of such an approach compared to the point dipoles is that it is better suited for simulation of polymer systems, as the movement of polar groups is restricted by the chain connectivity in the former model. The simulated system was rather straightforward – a single polyelectrolyte chain of the length N=256 in an explicit solvent. Each monomer unit of the chain consists of two sub-beads carrying charges $+q_1$ and $-q_2$, so the total charge of a unit is equal to $Q = q_1 - q_2$. To preserve the system electroneutrality, N=256 counterions were added to the system, the sub-beads of which carry point charges $-q_1$ and $+q_2$, thus resulting in the total charge $-Q$. Solvent beads are formed by two sub-beads with the charges $+q_3$ and $-q_3$, so they are uncharged ($Q_{solv} = 0$) and have the dipole moment of $p_s = q_3 d$. The bulk liquid dielectric permittivity values for different dipole moments $p_s$ were taken from the work[29]. Fig. 1 depicts the schematic representation of our setup.

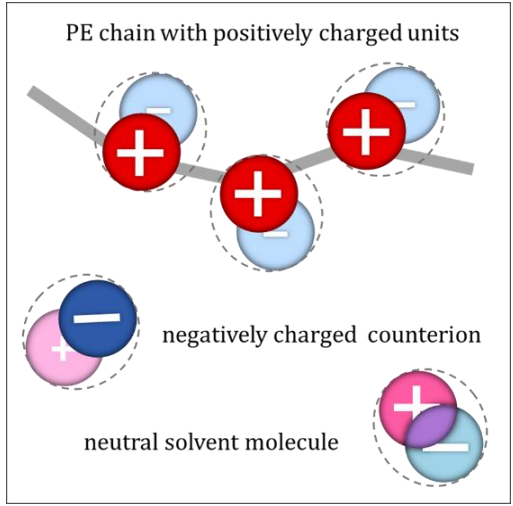

Fig.1 Schematic representation of the system under study.

The simulation box size was equal to $40^3$ and the bead number density was equal to 3. The soft-core interaction parameter between all the beads was equal to $a=100$, and the integration timestep was equal to dt=0.01. The bonded interactions were simulated using the harmonic potential with $=10$, $l_0=0.85$. Following the previous works on the topic,[6,27,31] we use the dimensionless parameter $\lambda^2 = \frac{e^2 Q^2}{\varepsilon k_b T R_c} = \frac{Q^2 l_b}{R_c} = \frac{\lambda_0^2}{\varepsilon}$ in order to characterize the strength of the electrostatic interactions, where $e$ is the elementary charge, $\varepsilon$ is the medium dielectric permittivity, $R_c$ is the soft-core cutoff radius (used as the length scale) and $l_b$ is the Bjerrum length. As it was mentioned earlier, in simulations one can change the value of $\varepsilon$ either by varying the "background" permittivity of the whole system (by simply adjusting the strength of all the electrostatic interactions) or explicitly by studying polar species.

First of all, let us compare these two approaches for taking into account the permittivity of the medium. Usually, in the works studying the aspects of the polyelectrolyte chain collapse, the difference in the dielectric permittivity between the polymer and the solvent is not taken into account;[6–8,31] therefore, we need the monomer units and even the counterions to have the same dielectric properties as the solvent beads to alleviate the possible influence of the dielectric mismatch (see below). To that end, for the monomer units and counterions we used the sub-bead charges of $|q_1| = Q + q_3$ and $|q_2| = q_3$ (i.e. the charge modules of the second sub-bead was equal to the charge modulus of the solvent sub-beads). One should note that while within the "background" permittivity approach changing $\varepsilon$ is completely equivalent to changing $\lambda_0^2$ (since both just rescale the strength of the electrostatic interactions between the charged monomer units and counterions), when considering explicit polar species it is not the case, since the dielectric effects are realized through the explicit interactions with large number of polar beads. Therefore, one can consider two separate approaches to changing $\lambda^2$: 1) varying $\lambda_0^2$ (by changing Q or kT which would result in the rescaling of the electrostatic interactions) at a fixed $\varepsilon$ or 2) varying $\varepsilon$ (by changing the dipole moment of the beads) at a fixed $\lambda_0^2$.

Let us consider the first case; the resulting dependence of the squared chain gyration radius $R_g^2$ on the electrostatic interaction strength $\lambda$ is presented in comparison to the results of homogeneous dielectric background approach (which were obtained by simulating the system with all the dipole moments equal to 0) in Fig.2, top.

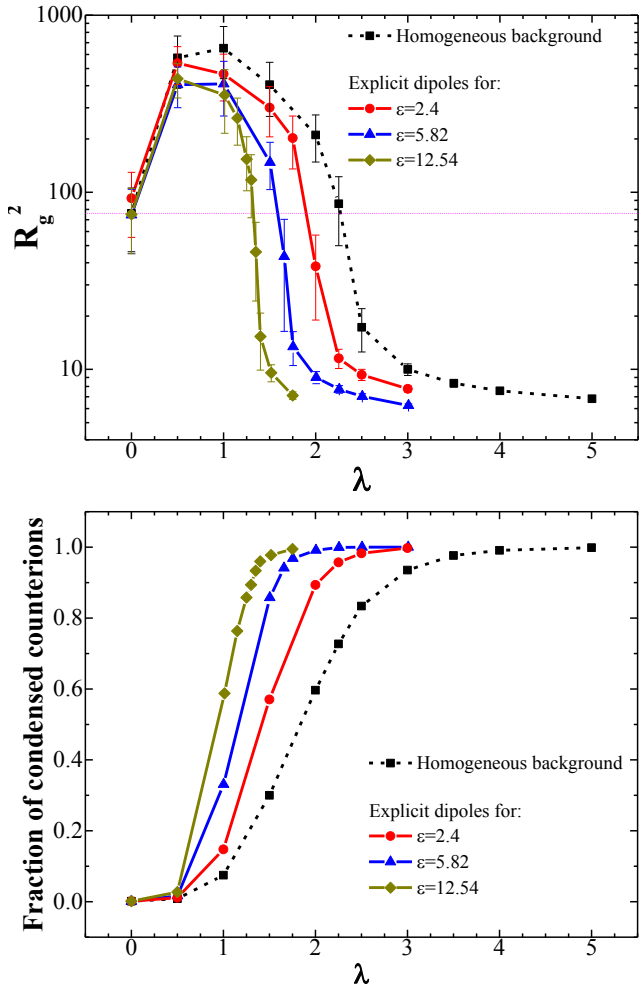

Fig.2. Top: dependences of $R_g^2$ on $\lambda$ for three different fixed values of $\varepsilon$ within the explicit polarity approach in comparison to the case of "background" permittivity approach. Bottom: dependence of the fraction of condensed counterions on $\lambda$ for three different fixed values of $\varepsilon$ within the explicit polarity approach in comparison to the case of "background" permittivity approach. A counterion was considered condensed if it was located closer than 0.6 from any monomer unit.

We observe an interesting feature – when the polarity of the species is simulated explicitly, the coil-to-globule transition occurs at smaller values of $\lambda$ compared to the case of the "background" permittivity approach. Moreover, we see that for $\varepsilon=2.4$ the collapse occurs at higher values of $\lambda$ compared to the case of $\varepsilon=5.82$, and for $\varepsilon=12.54$ the transition is shifted towards smaller values of $\lambda$ even further. One should keep in mind that, or course, for $\varepsilon=12.54$ much larger values of the total charge $Q$ (or lesser temperatures) are necessary for the transition than for $\varepsilon=2.4$, but in terms of the widely used electrostatic parameter $\lambda$ the transition still occurs earlier. The possible reasons for such behavior are discussed below. Let us now move to the second approach to changing $\lambda^2$ (varying $\varepsilon$ at a fixed $\lambda_0^2$); the dependence of $R_g^2$ on $\lambda$ is presented in Fig.3, top.

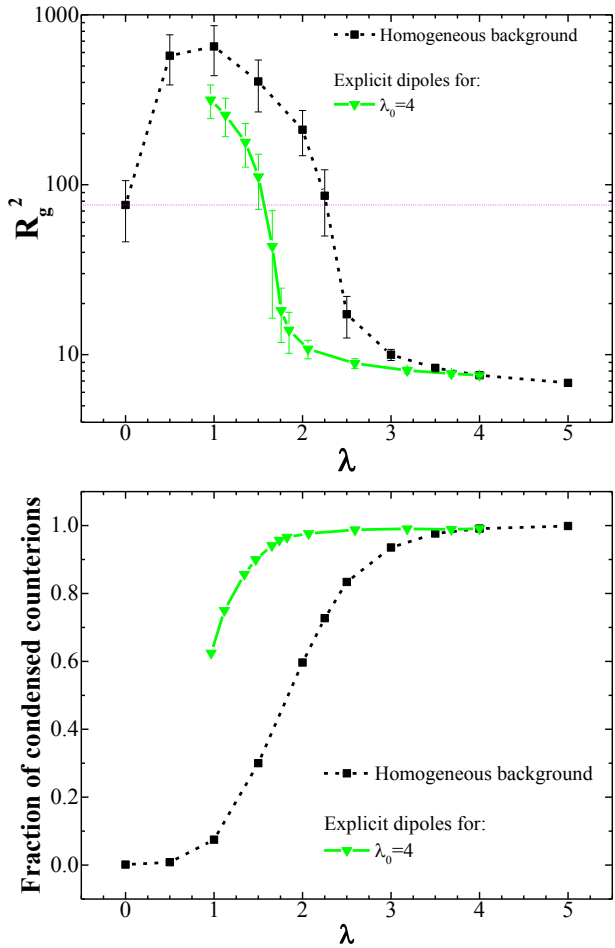

Fig.3 Top: dependence of $R_g^2$ on $\lambda$ for a fixed value of $\lambda_0$ (equal to 4) within the explicit polarity approach in comparison to the case of "background" permittivity approach. Bottom: dependence of the fraction of condensed counterions on $\lambda$ for a fixed value of $\lambda_0$ (equal to 4) within the explicit polarity approach in comparison to the case of "background" permittivity approach. A counterion was considered condensed if it was located closer than 0.6 from any monomer unit.

The same behavior is observed for this case as well. We believe that the reason of these peculiarities is that the mean-field notion of the dielectric permittivity is not entirely accurate when the distance between the charges becomes comparable to the solvent molecule size. Indeed, in the work [28,30] the interaction force between two charges in a polar liquid was studied, and it was shown that the effective dielectric permittivity (i.e. the magnitude of the force decrease upon placing the charges into the polar liquid) is actually distance-dependent. While at large distances it seems to converge to the bulk liquid permittivity, it fluctuates significantly when the charges come close to each other, and finally becomes much smaller that the bulk value when the charges are in close contact (i.e. form ion pair). We obtained the ion pair energy by integrating the interaction force, and such energy was found to be higher than for the case of the "background" permittivity approach. This circumstance leads to a higher faction of condensed counterions resulting in more compact chain conformations in the polyelectrolyte regime (i.e. when the counterions are released into the solution) at intermediate values of $\lambda \sim 0.5\text{-}1.5$; Fig.2, bottom, and Fig.3, bottom, depict the dependences of fraction of condensed counterions on $\lambda$. Moreover, when the chain starts to compact upon increasing the strength of electrostatic interactions $\lambda$, the charge-correlation attraction (which is believed to be responsible for the chain

collapse[4]) seem to be much stronger in the case of explicit polar species. Indeed, upon decreasing the chain volume the fraction of polar solvent beads (responsible for the screening of the electrostatic interactions) decreases, which intensifies the correlation attraction between the charged beads; in some sense, this resembles an "avalanche"-like effect, related to the presence of dielectric mismatch, described in the theoretical works on polyelectrolyte gels.[17] In the case of a compact globule observed at high enough $\lambda$, the entire globule volume consists of charged beads (monomer units + counterions), and the dielectric effect in this case is very weak (if present at all) even given the fact that every such charged bead is polar itself, since it is a collective effect for which sufficient amount of polar species is necessary. From a broader perspective, such behavior can in fact be viewed as a more general notion of dielectric mismatch between the solvent and the polymer chain volume (even though the monomer units and counterions have the same polarity as the solvent beads), realized through the intensification of the electrostatic interactions (i.e. decrease of the "effective" dielectric permittivity measured as the magnitude of the reduction of the electrostatic energy compared to the case of vacuum) inside the chain volume upon its collapse. From these observations we can draw a conclusion that the value of $\lambda$ seems to not have a universal meaning due to the small-scale effects related to the presence of polar species; therefore, changing the total unit charge Q or the temperature *kT* is not equivalent to changing $\varepsilon$.

So far we have studied the case of a polar backbone and "polar" counterions. In reality the polymer chains are usually much less polar (their typical permittivity lie in the range 2-8, see, for example[32]) than the majority of polar solvents, and the counterions are small molecules (or even atoms) having a zero permanent dipole moment. The dielectric contrasts of the constituent species have been shown to have a dramatic effect on the behavior of a number of polymer systems.[15–24] Let us now investigate whether the aforementioned "avalanche"-like effect predicted for the polyelectrolyte gels[17] is even more pronounced if the more realistic case of a non-polar chain is considered. To that end, the monomer units and counterions had their total charge placed only on one sub-bead, while the other was uncharged. Fig.4, top, depicts the obtained dependence of $R_g^2$ on $\lambda$ for the case of fixed value $\varepsilon=5.82$ in comparison with the case of polar chain and counterions, and Fig.4, bottom, presents the same comparison but for the case of fixed $\lambda_0=4$.

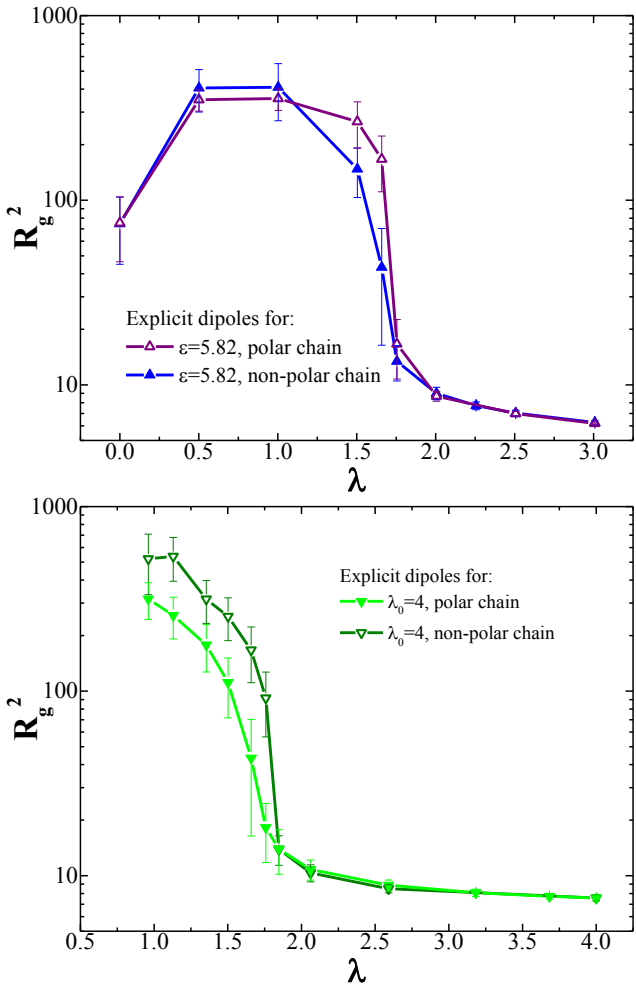

Fig.4 Top: obtained dependence of $R_g^2$ on $\lambda$ for the case of fixed value $\varepsilon=5.82$ in comparison with the case of polar chain and counterions; Bottom: obtained dependence of $R_g^2$ on $\lambda$ for the case of fixed value $\lambda_0=4$ in comparison with the case of polar chain and counterions.

Surprisingly, the non-polar chain collapses at larger values of $\lambda$, which is the opposite to what was expected; it also adopts somewhat more stretched conformation in the polyelectrolyte regime compared to the case of polar chain. These features are much more pronounced for the case of fixed $\lambda_0$ because the dipolar moment of the beads is rather large for the intermediate $\lambda$-values from 1 to ~2 in that case. The results may be explained by the presence of two factors: 1) the dipole-dipole interactions in the case of the polar chain result in an additional weak attraction leading to more compact chain conformation and earlier collapse; 2) for the charged beads it is more energetically favorable to be surrounded by polar beads, thus, the non-polar chain tends to adapt more swollen conformations to increase the number of contacts with polar solvent beads, which seems to be less pronounced in the case of the polar chain.

Summarizing, in this work we investigated the question of how the molecular nature of the dielectric media and the polymer-solvent dielectric mismatch affect the collapse of a polyelectrolyte chain in solution. To that end, we performed dissipative particle dynamics simulations with explicit treatment of polar species. First, we studied whether the explicit treatment of dielectric media as polar beads instead of homogeneous dielectric background results in a different system behavior. On an example of a polymer chain with the monomer units having the same polarity as the solvent beads, we showed that the explicit treatment of polar beads facilitates the chain collapse, i.e. it occurs at smaller values of the electrostatic strength parameter values λ. We believe that the main reason for such behavior is that the dielectric response is in fact a collective effect, and the "effective" dielectric permittivity (i.e. the magnitude of the reduction of the electrostatic energy) is different from the bulk value when the charges are close to each other and/or the density of the charges is high enough. Indeed, the ion pair energy is higher in the case of explicit polar species treatment compared to the case of homogeneous background even if the bulk permittivity is the same, leading to a higher fraction of condensed counterions; moreover, upon decreasing the chain volume the fraction of polar solvent beads (responsible for the screening of the electrostatic interactions) decreases, which intensifies the correlation attraction between the charged beads causing the chain to collapse easier. In some sense, such behavior can in fact be viewed as a more general notion of dielectric mismatch between the solvent and the polymer chain volume (even though the monomer units and counterions have the same polarity as the solvent beads), realized through the intensification of the electrostatic interactions (i.e. decrease of the "effective" dielectric permittivity) inside the chain volume upon its collapse. The electrostatic strength λ usually used as the main variable parameter in the works on the polyelectrolyte collapse seems to not have a universal meaning due to the small-scale effects related to the presence of polar species; therefore, changing the total unit charge Q or the temperature $kT$ is not equivalent to changing ε. Next, we investigated how the difference of the dielectric permittivities of the polymer chain and solvent affects the collapse. We showed that the polar chain adapts less swollen conformations in the polyelectrolyte regime and collapses easier compared to the case of non-polar chain. We believe that such surprising behavior can be explained by two reasons. First, the dipole-dipole interactions in the case of the polar chain result in an additional weak attraction leading to more compact chain conformation and earlier collapse; second, it is more energetically favorite for the charged beads to be surrounded by polar beads, thus, the non-polar chain tends to adapt more swollen conformations to increase the number of contacts with polar solvent beads. We hope that our findings will allow better understanding of the physics governing the behavior of polyelectrolytes in polar media.

## Acknowledgements

The financial support of the Russian Foundation for Basic Research (project 20-33-70164) is greatly acknowledged. The research is carried out using the equipment of the shared research facilities of HPC computing resources at Lomonosov Moscow State University.